\documentclass[a4paper,12pt]{article}
\usepackage{amsmath}
\usepackage{wasysym}
\usepackage{graphics}
\usepackage{graphicx}
\usepackage{epsfig}
\newcommand{\be} {\begin{equation}}
\newcommand{\ee} {\end{equation}}
\newcommand{\bma} {\begin{math}}
\newcommand{\ema} {\end{math}}
\newcommand{\beqa} {\begin{eqnarray}}
\newcommand{\eeqa} {\end{eqnarray}}
\newcommand{\al} {\alpha}
\newcommand{\ga} {\gamma}

\newcommand{\sig} {\sigma}
\newcommand{\nn} {\nonumber}
\newcommand{\bc} {\begin{center}}
\newcommand{\ec} {\end{center}}
\def\vec#1{\mathchoice{\mbox{\boldmath$\mathrm\displaystyle#1$}}
{\mbox{\boldmath$\mathrm\textstyle#1$}}
{\mbox{\boldmath$\mathrm\scriptstyle#1$}}
{\mbox{\boldmath$\mathrm\scriptscriptstyle#1$}}}
\newcommand{\bm}[1]{\mbox{\boldmath$#1$}}  
\renewcommand{\vec}{\bm}
\newcommand{\simgt}{\hbox{ \raise3pt\hbox to 0pt{$>$}
    \raise-3pt\hbox{$\sim$} }}
\newcommand{\simsm}{\hbox{ \raise3pt\hbox to 0pt{$<$}
    \raise-3pt\hbox{$\sim$} }}
\begin{document}
\hfill{HD-THEP-05-14}
\begin{center}
{\Large{\bf The Missing Odderon}}
\end{center}

\bigskip

\bc
{\large A Donnachie\footnote{sandy.donnachie@manchester.ac.uk}}\\
{\large School of Physics and Astronomy, University of Manchester}\\
{\large Manchester M13 9PL, England}
\ec

\bc
{\large H G Dosch\footnote{h.g.dosch@thphys.uni-heidelberg.de}, 
O Nachtmann\footnote{o.nachtmann@thphys.uni-heidelberg.de}}\\
{\large Institut f{\"u}r Theoretische Physik, Universit{\"a}t Heidelberg}\\
{\large Philosophenweg 16, 69120 Heidelberg, Germany}
\ec

\begin{abstract}
\noindent In contrast to theoretical expectations, experimental results at 
$\sqrt{s}=200$ GeV for the reaction $\gamma p \to \pi^0 X$ show no evidence 
for odderon exchange. The upper limit on the cross section is an order of 
magnitude smaller than the theoretical estimate. It is argued that chiral 
symmetry leads to a large suppression, taking the theoretical estimates well 
below the data. Two additional arguments are presented which may decrease the 
theoretical estimate further. The calculations are more sensitive to the 
assumptions made in evaluating the hadronic scattering amplitude than in the 
processes considered previously and lattice gauge calculations indicate that 
the odderon intercept may be appreciably lower than usually assumed. These 
two latter effects are particularly relevant for the reactions $\gamma p \to 
f_2^0(1270) X$ and $\gamma p \to a_2^0(1320) X$ for which the data upper 
limits are also below the theoretical predictions, but not so dramatically 
as for $\gamma p \to \pi^0 X$.
\end{abstract}

\section{Introduction}

The phenomenological pomeron has long been established as an effective
Regge pole with trajectory $\al_{\rm pom} \approx 1.08 + 0.25t$ whose exchange
governs high-energy diffractive scattering \cite{DDLN02}. There is no 
{\it a priori} reason why the phenomenological odderon, a $C = P = -1$ 
partner of the $C = P = +1$ pomeron, should not exist \cite{Nic_1}. Indeed
within perturbative QCD, the odderon is rather well defined with an 
intercept $\al_{\rm odd}(0) \approx 1$, see \cite{odd_p}. For a general review 
of odderon physics see \cite{Ewerz}. Applications in
the nonperturbative regime \cite{Nic_2} have assumed a ``maximal''
odderon with an intercept $\al_{\rm odd}(0) = 1$. The exchange of the
phenomenological odderon should produce a difference between $p p$ and 
$\bar{p}p$ scattering at high energy and small momentum transfer, a
particularly sensitive test being provided by the forward real part of 
the  $p p$ and $\bar{p}p$ scattering amplitudes. However, measurements 
\cite{odd_exp} are consistent with the absence of odderon exchange. 
An explanation is provided \cite{svm_1} by the clustering of two quarks to 
form a small diquark in the nucleon which has the effect of suppressing the
odderon-N-N coupling and completely so for a pointlike diquark. 
However if one (or both) of the nucleons is
transformed into an excited negative-parity state then the odderon
can couple without any restriction according to \cite{svm_1}. In contrast 
to the apparently ``missing odderon'' at very small momentum transfers there 
is experimental evidence for $C=-1$ exchange at larger momentum transfers. 
The $pp$ and $\bar{p}p$ differential cross sections \cite{100a} differ 
markedly for $|t|\approx 1.3$ GeV$^2$ in the ISR energy range. For still 
larger $|t|$ the $C=-1$ exchange is even supposed to dominate. We shall 
discuss these two points in section 4 below. 

As an alternative to $p p$ and $\bar{p}p$ scattering it was suggested
\cite{HERA,8a} that high-energy photoproduction of $C=+$ mesons, e.g. 
$\pi^0$, $f_2^0(1270)$ and $a_2^0(1320)$, with nucleon excitation
would provide a clean signature for odderon exchange. Specific calculation 
\cite{Heid_1,Heid_2}
predicted the following cross sections at $\sqrt{s} = 20$ GeV:
\beqa \label{1}
\sig(\ga p \to \pi^0 X) &\approx& 300~~{\rm nb}\nn\\
\sig(\ga p \to f_2^0(1270) X) &\approx& 21~~{\rm nb}\nn\\
\sig(\ga p \to a_2^0(1320)) X) &\approx& 190~~{\rm nb}
\label{predict}
\eeqa
The experimental results at $\sqrt{s} = 200$ GeV for $\pi^0$ \cite{H1a}, 
$f_2(1270)$ and $a_2(1320)$ \cite{H1b} are:
\beqa\label{2}
\sigma(\ga p \to \pi^0 N^*) &<&  49~~{\rm nb}\nn\\
\sigma(\ga p \to f_2^0(1270) X) &<& 16~~{\rm nb}\nn\\
\sigma(\ga p \to a_2^0(1320) X) &<& 96~~{\rm nb}
\label{expt}
\eeqa
all at the 95$\%$ confidence level.
The model was based on an approach to high-energy diffractive scattering
using functional integral techniques \cite{Nac91} and an extension 
\cite{svm_2} of the model of the stochastic vacuum \cite{svm_3}. This model gives a 
remarkably good
description of many different processes dominated by pomeron exchange
\cite{svm_2,Heid_4}. It is easily extended to odderon exchange and gives an odderon intercept $\al_{\rm odd}(0) = 1$. The scattering amplitude 
$T(s,t)$ is obtained through a profile function $J(\vec b,s)$ :
\be
T(s,t) = 2is \int \, d^2b\;{\rm exp}(i\vec{q}\cdot\vec{b})\,J(\vec b, s).
\end{equation}
The function $J(\vec b,s)$ is given in turn by the overlap of a dipole-dipole 
scattering amplitude $\tilde{J}(\vec b, \vec r_1, \vec r_2, z_1, z_2)$
with appropriate wave functions for the initial and final states:
\beqa
J(\vec b, s) &=&-\int \frac{d^2r_1}{4\pi}dz_1\int \frac{d^2 r_2}{4\pi}dz_2\nn\\
&&\sum
\Psi^*_M(\vec{r}_1,z_1)\Psi_{\ga}(\vec{r}_1,z_1)
\Psi^*_{p^\prime}(\vec{r}_2,z_2)\Psi_{p}(\vec r_2,z_2)
\tilde{J}(\vec b, \vec r_1, \vec r_2, z_1, z_2).
\eeqa
Here $\vec b$ is the impact parameter of two light-like dipole trajectories 
with transverse sizes $\vec r_1$ and $\vec r_2$ respectively and $z_1$, $z_2$ 
are the longitudinal momentum fractions of the quarks in the dipoles. The
physical picture is that the photon fluctuates into a $q\bar q$ pair, this
is turned into the final meson $M$ by the soft colour interaction $\tilde{J}$, 
determined from other reactions \cite{Heid_4} and the proton is excited
into an appropriate baryon resonance. The nucleon 
and the baryon resonances are treated as quark-diquark dipole systems. The 
wave functions automatically take 
into account helicity flip at the particle and at the quark level and 
produce the correct helicity dependence of $d\sig/dt$ as $t \to 0$ for
Regge-pole exchange. The cross sections for $\pi^0$, $f_2^0(1270)$ and 
$a_2^0(1320)$ photoproduction were evaluated at $\sqrt{s} = 20$ GeV as
that is the energy at which the parameters of $\tilde{J}(\vec b, \vec r_1, 
\vec r_2, z_1, z_2)$ were obtained. In elastic hadron-hadron scattering the 
increase of the cross sections, together with the shrinking of the 
diffractive peak, can be reproduced in this model by suitable scaling of 
the hadronic radii. The assumption that the same radial scaling is relevant
for the energy dependence of the odderon contributions, leads to the
photoproduction cross sections scaling as $(\sqrt{s}/20)^{0.3}$ and
to an enhancement of about 2 at $\sqrt{s}=200$ GeV.

The results for diffractive dissociation depend much more on the choice of
wave functions than for elastic processes. In the latter the overlap is
essentially the density and is constrained by the normalisation, which is
not the case for the former. The photon-meson overlap is tested to some 
extent by the known radiative decays of the meson, but there is no such 
test for the overlap between the proton and the final baryonic state. Also,
the odderon exchange is much more sensitive to the parameters of the model
than is pomeron exchange, see section 3 below. For these reasons it was 
suggested \cite{Heid_1,Heid_2} that the uncertainty in the model calculation 
is at least a factor 2 at $\sqrt{s} = 20$ GeV.

The results (\ref{expt}) are well below the predictions (\ref{1}), for the 
$\pi^0$ drastically so. Thus, from $pp$ and $\bar{p}p$ scattering as well as 
from meson production one concludes that the odderon is apparently ``missing'' 
at small $|t|$. It is important to
understand why this may be so. In the following we reconsider each part of the
calculations of \cite{Heid_1,Heid_2}.

\section{Wave functions}

Consider first $\pi^0$ photo- and electroproduction at high energies
\be\label{2.1}
\gamma^{(*)}(q)+p(p)\rightarrow \pi^0(q')+X(p')
\end{equation}
where $X$ is a proton or a diffractively-excited proton state. We shall argue 
that chiral 
$SU(2)\times SU(2)$ symmetry of QCD, which is broken only by the small $u$ and 
$d$ quark masses, 
leads to a large suppression factor for this reaction. A detailed account of 
this will be given 
in \cite{100}. Here we only outline the arguments. 

\begin{figure}[t]
\begin{center}
\begin{minipage}{85mm}
\epsfxsize85mm
\epsffile{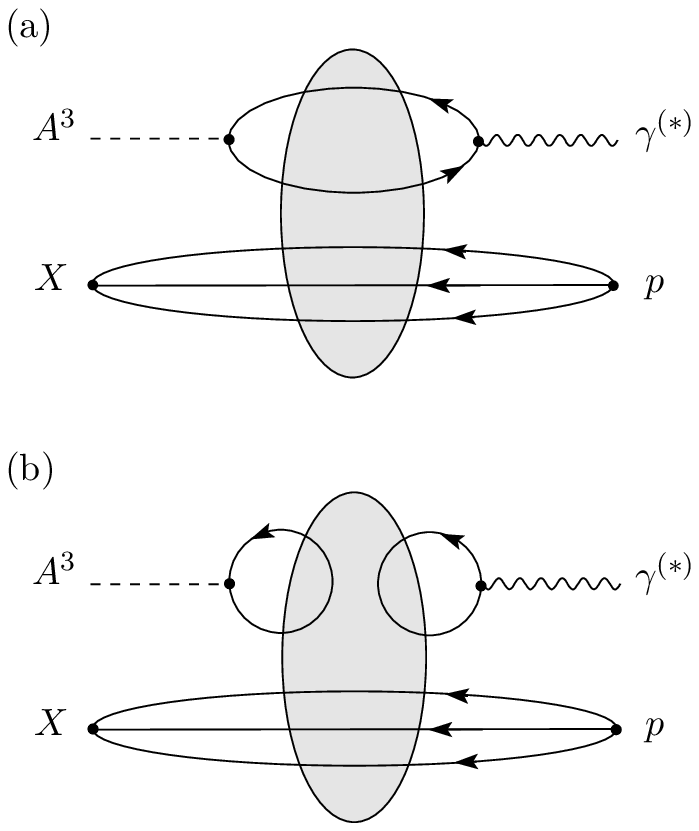}
\end{minipage}
\end{center}
\caption{Diagrams of type (a) and (b) for the process $\gamma^{(*)} p
\rightarrow A^3X$. The full 
lines correspond to quark propagators in a given gluon potential. The shaded 
blobs indicate the 
functional integration over all gluon potentials with a measure including the 
fermion determinant. For the process $\gamma^{(*)} p\rightarrow \gamma^{(*)} p$
replace $A^3$ by $\gamma^{(*)}$ and $X$ by $p$.}
\label{Fig.1}
\end{figure}

Real and virtual Compton scattering
\be\label{2.2}
\gamma^{(*)}(q)+p(p)\rightarrow \gamma^{(*)}(q')+p(p')
\end{equation}
were investigated in \cite{101,102} using exact functional techniques. A 
classification of diagrams into seven types 
(a) to (g) (see Fig.2 of \cite{101}) was given. The diagrams of types (a) and 
(b), which are 
the relevant ones for our discussion here, are as shown in Fig.1 of the present
paper with $A^3$ replaced by 
$\gamma^{(*)}$ and $X$ by $p$. The diagram types are distinguished by the 
topology of the quark loops and the placing of the photon coupling on them. 
It was argued in \cite{101,102} that at high energies the diagrams of types 
(a) and (b) are the leading ones. 

Now one can use PCAC (partial conservation of the axial current) to relate 
reaction (\ref{2.1}) to a process very similar to (\ref{2.2}) 
where the final state $\gamma^{(*)}$ is replaced by the third isospin 
component of the axial current, $A^3_\mu$, and the final state proton by $X$
\be\label{2.3}
\gamma^{(*)}(q)+p(p)\rightarrow A^3(q')+X(p').
\end{equation}
The isotriplet of axial currents is 
\be\label{II.4}
A^a_\mu(x)=\bar{q}(x)\gamma_\mu\gamma_5\frac{\tau^a}{2}q(x),
\end{equation}
where $a=1,2,3,~\tau^a$ are the Pauli matrices and 
\be\label{II.5}
q(x)=\left(
\begin{array}{c}
u(x)\\d(x)\end{array}\right)
\end{equation}
is the quark field operator. The well-known PCAC relation is
\be\label{II.6}
\partial_\lambda A^{a\lambda}(x)=\frac{f_\pi m^2_\pi}{\sqrt{2}}
\phi^a(x),
\end{equation}
where $\phi^a(x)$ is a correctly normalised pion field and $f_\pi ~
\widetilde{=}~0.93 m_\pi$ is 
the pion decay constant. By PCAC the amplitudes of the reactions (\ref{2.1}) 
and (\ref{2.3}) are related by
\be\label{II.7}
iq'_\mu {\cal M}^{\mu\nu}(A^3;q',p,q)=
-\frac{f_\pi m^2_\pi}{2\pi m_p\sqrt{2}}\frac{1}{q^{'2}-m^2_\pi+i\epsilon}
{\cal M}^\nu(\pi^0;q',p,q).
\end{equation}
Here $m_p$ is the proton mass and we extrapolate the amplitude for (\ref{2.1}) 
from on shell 
pions, $q^{\prime 2}=m^2_\pi$, to arbitrary $q^{\prime 2}\leq m^2_\pi$. One 
can then show the following.
\begin{itemize}
\item At high energies the diagrams of types (a) and (b) shown in Fig.1 are 
the leading ones for reaction (\ref{2.3}). 
\end{itemize}

For simplicity we discuss in this note only the isospin-symmetry limit, that 
is we set for the current quark masses
\be\label{2.4}
m_u=m_d \equiv\hat{m}.
\end{equation}
The quark loop attached to the current $A^3_\mu$ in Fig.1b must then vanish, 
since $\tau^3$ is the only nontrivial flavour matrix in this loop and 
tr$\{\tau^3\}=0$. Thus we find
\begin{itemize}
\item In the isospin-symmetry limit the diagrams of type (b) vanish for the 
reaction (\ref{2.3}) and, using (\ref{II.7}), also for pion production 
(\ref{2.1}). 
\end{itemize}

A more involved analysis is necessary for the diagrams of type (a). Using PCAC 
(\ref{II.6}) one can show the following (see \cite{100}):
\begin{itemize}
\item The diagrams of type (a) when inserted in (\ref{II.7}) give a 
contribution to the $\pi^0$ amplitude which is proportional to $\hat{m}$, that 
is to the current quark mass. 
\end{itemize} 
The current quark mass is proportional to $m^2_\pi$ in the chiral limit (see 
for instance eq. (8.1) of \cite{103})
\be\label{2.5}
m^2_\pi=2\hat{m}B~,~B=-\frac{2}{f^2_\pi} \langle 0|\bar{u}u|0\rangle.
\end{equation}
Typical values for $\hat{m}$ and $B$ at a renormalisation scale of 1 GeV are 
$\hat{m}=7$ MeV, $B=1.4$ GeV. 

Thus we find that the amplitude for (\ref{2.1}) is proportional to $m^2_\pi$, 
that is it vanishes in the chiral limit. To estimate the actual suppression 
factor $\kappa$ in the amplitude relative 
to a naive estimate, such as the one given in \cite{Heid_1}, we argue as 
follows. To get a 
dimensionless factor we divide $m^2_\pi$ by a typical hadronic squared mass 
scale, say $m^2_p$, and write 
\be\label{II.10}
\kappa=\frac{m^2_\pi}{m^2_p}h.
\end{equation}
Here $h$ should be of order 1 but could be numerically large, for instance 
$h=m_p/f_\pi~\widetilde
{=}~7$. Putting everything together we estimate for the suppression factor 
\beqa\label{II.2}
0.02 \apprle\kappa\apprle 0.15,\nonumber\\
5\times 10^{-4}\apprle\kappa^2\apprle0.02.\nonumber\\
\eeqa
That is, the cross section for $\gamma p\rightarrow \pi^0 X$ should be 
suppressed at least by a factor $\approx 50$. 

Thus one shortcoming of the calculation \cite{Heid_1} for $\gamma p\rightarrow
\pi^0X$ is that the $\pi^0$ wave function used did 
not properly take into account the constraints from chiral symmetry. This 
clearly reduces the theoretical prediction for 
(\ref{2.1}). Of course, other effects, as discussed in this note, may reduce 
the theoretical estimate further. 

\begin{figure}[t!]
\begin{center}
\begin{minipage}{85mm}
\epsfxsize85mm
\epsffile{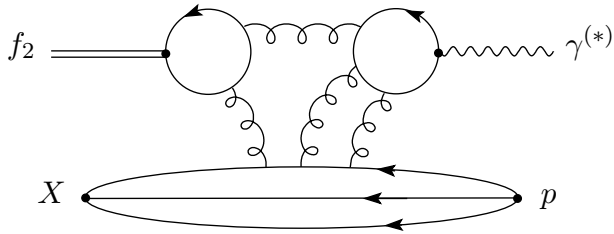}
\end{minipage}
\end{center}
\caption{A specific diagram of type (b) for $\gamma^{(*)}p\rightarrow 
f_2^0 X$.}
\label{Fig.2}
\end{figure}

What can we say about $f_2^0$ and $a_2^0$ production? We note that the 
$f^0_2(1270)$ is an isoscalar, the $a^0_2(1320)$ an isovector particle. 
One can again make a general analysis in terms of diagram types, replacing 
$A^3$ in Fig.1 by appropriate tensor currents. Here we only note 
that in the isospin-symmetry limit the same arguments as given for the 
$A^3_\mu$ case above show that diagrams of type (b) cannot contribute to 
$\gamma^{(*)}p\rightarrow a_2X$ but can contribute to 
$\gamma^{(*)}p\rightarrow f_2^0 X$. A simple diagram of this type is shown 
in Fig.2. Thus, if for some reason diagrams of type (a) in Fig.1 are 
suppressed also for tensor meson production, then $a_2^0$ will be suppressed 
but $f_2^0$ need not be. Data \cite{H1b} may give a hint in this direction. 

\section{Soft-colour interaction}

The functional approach to quantum field theory has turned out to
be a most effective one for investigating nonperturbative effects in
QCD. In it the expectation values of field operators are expressed
as functional integrals over the classical fields, where the
weight of the configuration is given by the exponential of the QCD
action. This functional integration takes into account the quantum
fluctuations. The short-range fluctuations can be calculated with
the help of perturbation theory, but for a treatment of the
effects of long-range fluctuations numerical simulations or model
assumptions are necessary.

The stochastic vacuum model (for a review see \cite{DDSS02}) is one
such approach to nonperturbative QCD. It assumes that the long-range 
fluctuations can be approximated by the only functional
integral which is analytically accessible, namely a Gaussian
functional integral (Gaussian process). The Gaussian approximation
is defined through the cumulant or linked-cluster expansion 
\cite{VKa76} of the expectation value of several fields. If all
cumulants containing more than two fields are neglected we are
left with Gaussian integrals and all expectation values can then
be expressed through products of the expectation value of two
fields, the so-called correlator.

In a theory where the variables of the functional integration,
that is the classical fields, commute, the Gaussian approximation
is uniquely defined. In a non-Abelian field theory this is not the
case, since there are several cumulant expansions possible and the
truncation of them leads to different Gaussian approximations. An
additional complication is induced by the dependence on the path
connecting the space-time points of the two fields of the
correlator. This path has to be introduced in order to ensure
gauge independence.

For the investigation of the forces between two quarks and in
particular for confinement, the expectation value of a single
Wilson loop has to be calculated. In that case the so-called van
Kampen cumulants are a natural choice for the expansion and the
truncation of it leads to a Gaussian integral. This allows the
approximate evaluation of expectation values of a single Wilson
loop. This choice, together with the assumption that the paths
mentioned above have no influence on the correlator, led to several
highly desirable results \cite{DDSS02}. In particular:

\begin{itemize}
\item A non-Abelian gauge theory like QCD shows confinement. In order to 
obtain confinement in an Abelian gauge theory monopole
condensation has to occur.

\item When lattice results for the fundamental correlator are inserted, the
string tension comes out to have the correct phenomenological value.
Furthermore it is proportional to the Casimir operator of the
representation of the Wilson loop, a result which is also in 
good agreement with lattice calculations. 
\end{itemize}

These results support strongly the Gaussian approximation. Also the 
relativistic spin- and velocity-dependent terms of the interquark 
potential, as obtained from the model, are in agreement with phenomenology.

In order to evaluate hadronic scattering amplitudes the
expectation value of at least two Wilson loops has to be
calculated \cite{svm_2}; this follows from the formalism developed 
in \cite{Nac91}. However the functional
integration variables in the expansion used for the evaluation of
one loop cannot be used in that case. Therefore two new
different cumulant expansions  have been used, a simple expansion
method and a more sophisticated super-cumulant method, as
explained in detail in chapter 8.5 of \cite{DDLN02}. Both these cumulant
expansions differ from the one used for the evaluation of a single loop. Hence
the Gaussian approximations made for the single loop and that made for 
scattering processes are not the same. The details are discussed in 
\cite{Rd95,Nac97,DDSS02}.

Nevertheless
the modified model with the same input parameters for the correlator 
gives very satisfactory results for scattering and production processes 
where pomeron exchange is dominant. With only a few parameters a whole 
range of experimental results could be reproduced and even predicted, 
see \cite{DDLN02,Heid_4}. The model can also be applied to processes which can 
only occur via the exchange of a $C$-odd state, for instance the 
photoproduction of neutral pions, and is at the core of the predictions of 
\cite{Heid_1,Heid_2}. The fact that the predicted cross section for 
$\ga p \rightarrow \pi^0X$ is an order
of magnitude larger than the upper limit of the experimental cross section 
for this reaction forces us to consider possible sources of error in the 
extended stochastic vacuum model. It was already mentioned that the underlying 
cumulant expansions used for the evaluation of scattering amplitudes are 
different from those used for the evaluation of one loop. For definiteness we 
call the cumulants of the latter W-cumulants and those of the former
S-cumulants (for scattering). Only the correlator (the cumulant of two fields) 
is the same in all expansions. The vanishing of the higher W-cumulants does  
not imply the vanishing of the higher S-cumulants.

Scattering processes with pomeron ($C$-even) exchange are
dominated by the expectation value of a product of four gluon
fields that is reduced to the product of two correlators. The
success  of the model for these processes shows that the cumulant
of four fields is indeed not only small for the W- but also  for
the S-expansions. For $C$-odd induced processes the leading term
is the expectation value of a product of six fields. In the model
it is factorised to a product of three correlators. This
factorisation is only justified if additionally the S-cumulant
for six fields is small.  It should be noted that the two
S-expansions mentioned above do not differ in the cumulant of four
fields but do in that of six fields, hence the S-cumulant of six
fields cannot vanish in both S-expansions. Thus a possible explanation 
for the discrepancy between the theoretical expectation (\ref{1}) and the 
experiment (\ref{2}) is a large S-cumulant of six fields compensating the 
product of the three correlators to a large extent. This is independent 
from the wave function effects discussed in sect. 2.

Lattice calculations could provide a test for this hypothesis of a
large S-cumulant of six fields \cite{SSDP03}. The leading term for
the difference between expectation values of a product of parallel
and antiparallel Wilson loops is the expectation value of six
fields. Comparison between model and full lattice calculations can
therefore test directly the factorisation hypothesis without
involving the folding with hadron wave functions which always
occurs in the evaluation of scattering or production processes.

\section{Energy dependence}

There is no {\it priori} justification for the assumption that the odderon 
trajectory should match the pomeron trajectory nor, in particular, that their 
contributions to elastic processes should have the same energy dependence. 
There is some evidence that they may indeed be different.    
The differential cross sections of elastic $p p$ and $\bar{p} p$ scattering
at $\sqrt{s} = 53$ GeV are different in the region of $|t| = 1.3$ GeV$^2$,
the $p p$ data having a marked dip which is not present in the $\bar{p} p$
data. This difference must be due to $C = -1$ exchange and is the only real 
experimental evidence for the existence of the odderon. The data can be 
fitted by including, in addition to the usual pomeron and other Regge 
poles, the maximal odderon \cite{Nic_2} or 3-gluon exchange 
\cite{DL_1} which is closer in spirit to the perturbative odderon than to 
the nonperturbative odderon of \cite{Nic_2} or the nonperturbative odderon of 
\cite{26a}. Indeed, in \cite{26a} the odderon contributions to $pp$ and 
$\bar{p}p$ scattering were calculated with the same methods discussed above 
for single-meson photoproduction. The proton was still considered as a 
quark-diquark system but now the diquark was assumed to have a radius $R$. 
It was shown in \cite{26a} that for $R=0$ the odderon effects in $pp$ versus 
$\bar{p}p$ at $|t|\approx 1.3$ GeV$^2$ vanish in accordance with \cite{svm_1} 
and that the data \cite{100a} could be reproduced with a -- very reasonable -- 
diquark radius $R\approx 0.22$ fm.

The $p p$ data at large $|t|$ appear to be 
essentially energy independent for $27.4 \leq \sqrt{s} \leq 62.1$ GeV and, 
in both models \cite{Nic_2} and \cite{DL_1}, are primarily odderon exchange. 
If the odderon is considered 
as a Regge pole then the near-constancy of the $p p$ large-$|t|$ cross section 
requires the maximal odderon to have a very flat trajectory. See \cite{Ewerz} 
for a full discussion of these points.

However, 
a rather different picture emerges from Lattice Gauge Theory \cite{Morn,Teper}.
In \cite{Teper}, the lightest $J=0,2,4,6$ glueball masses have been calculated
in the D$=3+1~SU(3)$ gauge theory and extrapolated to the continuum limit. 
Assuming that the masses lie on linear Regge trajectories, the leading 
glueball trajectory is found to be $\al(t)=(0.93 \pm 0.024)+(0.28 \pm 0.02)
\al^\prime_R t$, where $\al^\prime_R \approx 0.9$ is the slope of the usual 
mesonic Regge trajectories. Thus this glueball trajectory has an intercept 
and slope very similar to that of the pomeron trajectory, $\al_{\rm pom} 
\approx 1.08 + 0.25t$ \cite{DDLN02}. The states one might expect to lie on 
the odderon trajectory are the lightest $J^{PC}=1^{--},3^{--},5^{--},\cdots$. 
The lattice results for $1^{--}$ and $3^{--}$ define a trajectory with a 
slope similar to the pomeron but with a very low, negative intercept. These 
results are shown in figure 3. A similar conclusion about the 
odderon trajectory is reached in \cite{SK00} but from a very different 
standpoint. As the glueballs on the pomeron trajectory are two-gluon states 
and those on the odderon trajectory are three-gluon states, that the latter 
is low-lying is not surprising in a constituent-gluon picture as the 
effective gluon mass $\sim 1$ GeV. If this is the correct interpretation 
of the lattice calculations, then it completely destroys the odderon 
exchange model used to calculate $\pi^0$, $f_2^0(1270)$ and $a_2^0(1320)$ 
photoproduction.

\begin{figure}
\begin{center}
\begin{minipage}{90mm}
\epsfxsize90mm
\epsffile{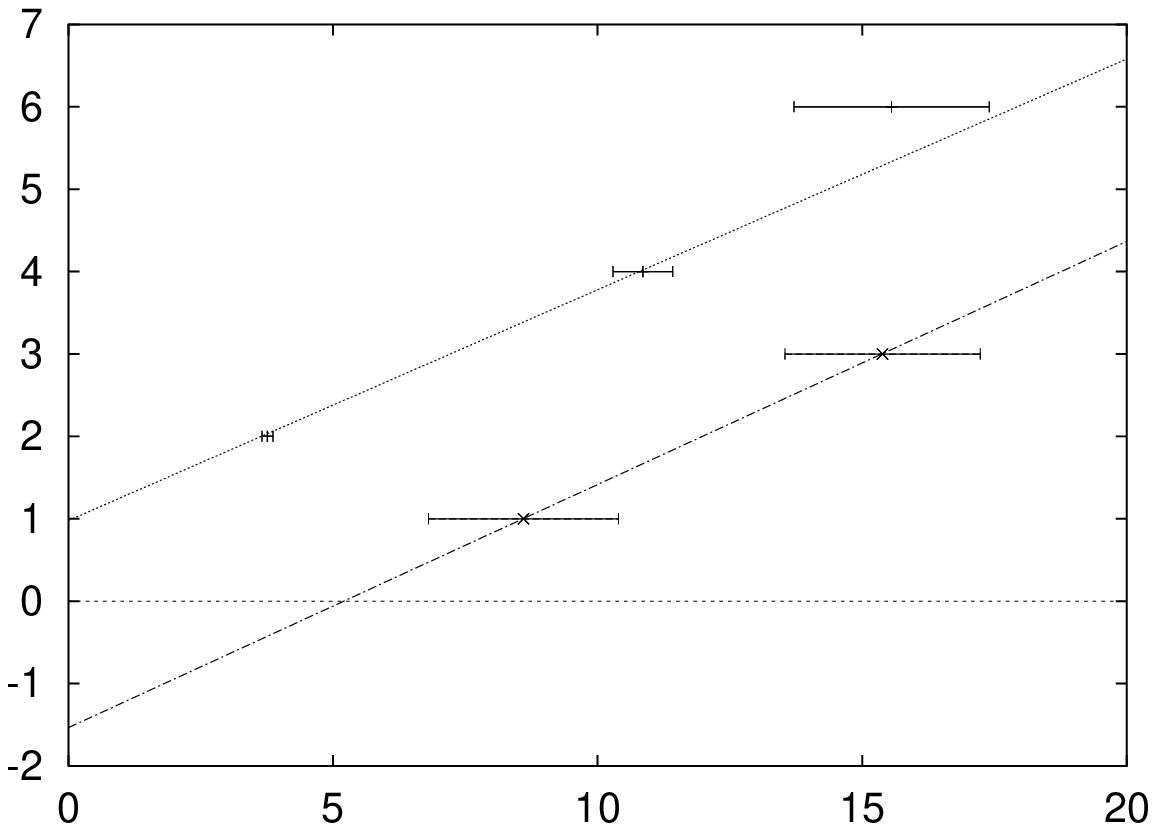}
\begin{picture}(0,0)
\setlength{\unitlength}{1mm}
\put(-5,45){\small{J}}
\put(40,-2){\small{$t$ (GeV$^2$)}}
\end{picture}
\end{minipage}
\caption{Pomeron and odderon trajectories from lattice gauge theory. The
data points are from \cite{Teper}.}
\end{center}
\label{traj}
\end{figure}

However there is an alternative explanation of the lattice result for the 
odderon. If the leading trajectory has an intercept around unity then the 
lightest $1^{--}$ glueball cannot lie on it but will lie on a subleading
trajectory. Drawing a linear trajectory from $J=1$ at $t=0$ through the 
mass of the lightest $3^{--}$ glueball gives a slope about half the slope 
of the pomeron trajectory. The ambiguity would be removed if the mass of the 
lightest $5^{--}$ could be calculated. This alternative explanation does 
not have a significant effect on the calculation of \cite{Heid_1,Heid_2}.

\section{Conclusions} 

We have presented three arguments highlighting aspects of the predictions
\cite{Heid_1,Heid_2} for odderon exchange in photoproduction which may have
been too optimistic. These are the role of wave functions and in particular
the effect of chiral $SU(2) \times SU(2)$ symmetry, the possible breakdown 
of the factorisation procedures used in calculating the interaction and 
uncertainties in the energy dependence of odderon exchange. The first of 
these leads to a large suppression for the reaction $\gamma^{(*)} p \to 
\pi^0 X$ and can account for the large discrepancy between the data \cite{H1a} 
and the original predictions. The other two aspects of the calculation are 
less quantifiable, but could provide an explanation of the discrepancies 
between prediction and experiment in the reactions $\gamma p \to f_2^0(1270) X$
and $\gamma p \to a_2(1320) X$. Both the factorisation procedure and the 
assumption on energy dependence are amenable to being checked by lattice-gauge 
calculations.

\bigskip
 
{\Large \bf Acknowledgements}

We are grateful to C Ewerz for providing figures
1 and 2. We have greatly profited from comments on the manuscript by C~Ewerz, 
H~Ch~Schultz-Coulon, and A~Utermann. 
Discussions with K~H~Meier, H~Ch~Schultz-Coulon and J~Stiewe  are gratefully 
acknowledged.
This work was supported in part by the German 
Bundesministerium f\"ur Bildung und Forschung under project no. HD 05HT4VHA/0.

\end{document}